\documentclass[aps,prl,twocolumn,superscriptaddress,showpacs,amsmath,amssymb]{revtex4-1}

\usepackage{graphicx}
\usepackage{dcolumn}
\usepackage{bm}
\usepackage{epstopdf}

\newcommand{\mr}[1]{\mathrm{#1}}
\newcommand{\be}{\begin{equation}}
\newcommand{\ee}{\end{equation}}

\newcommand{\figta}{$\left(\mathrm{a}\right)\;$}
\newcommand{\figtb}{$\left(\mathrm{b}\right)\;$}
\newcommand{\figtc}{$\left(\mathrm{c}\right)\;$}
\newcommand{\figtd}{$\left(\mathrm{d}\right)\;$}

\newcommand{\figa}{$\left(\mathrm{a}\right)$}

\newcommand{\figc}{$\left(\mathrm{c}\right)$}

\newcommand{\Mohm}{\;\mr{M}\Omega}
\newcommand{\kohm}{\;\mr{k}\Omega}

\newcommand{\mhz}{\;\mr{MHz}}
\newcommand{\ghz}{\;\mr{GHz}}

\newcommand{\mk}{\;\mr{mK}}

\newcommand{\nh}{\;\mr{nH}}
\newcommand{\na}{\;\mr{nA}}
\newcommand{\ns}{\;\mr{ns}}

\newcommand{\mm}{\;\mr{mm}}
\newcommand{\fm}{\;\mr{F}/\mr{m}}
\renewcommand{\hm}{\;\mr{H}/{\mr{m}}}
\newcommand{\ms}{\;\mr{m}/\mr{s}}

\newcommand{\pers}{\;\mr{s}^{-1}}

\newcommand{\mum}{\;\mu\mr{m}}

\newcommand{\nm}{\;\mr{nm}}

\newcommand{\bext}{B_{\mr{ext}}}
\newcommand{\phiext}{\Phi_{\mr{ext}}}

\newcommand{\phio}{\Phi_0}
\newcommand{\wq}{\omega_{\mr{q}}}

\newcommand{\fq}{f_{\mr{q}}}
\newcommand{\fp}{f_{\mr{p}}}
\newcommand{\fs}{f_{\mr{s}}}

\newcommand{\ip}{I_\mr{p}}

\newcommand{\lr}{L_\mr{r}}
\newcommand{\aq}{A_\mr{q}}
\newcommand{\ar}{A_\mr{r}}

\newcommand{\dphi}{\delta\Phi_{\mr{ext}}}
\newcommand{\clen}{C_l}
\newcommand{\llen}{L_l}
\newcommand{\vph}{v}

\newcommand{\deltaff}{\delta f_{\mr{FWHM}}}
\renewcommand{\lq}{L_{\mr{q}}}

\newcommand{\lsq}{L_{\square}}
\newcommand{\rrq}{R_{\mr{q}}}
\newcommand{\rsq}{R_{\square}}

\newcommand{\phip}{\varphi_{\mr{p}}}

\newcommand{\inox}{\mr{InO}_{\mr{x}}}
\newcommand{\pps}{P_{\mr{s}}}
\newcommand{\ppp}{P_{\mr{p}}}

\begin{document}

\title{Coherent dynamics and decoherence in a superconducting weak link}

\author{J. T. Peltonen}
\email{joonas.peltonen@riken.jp}
\altaffiliation[Present Address: ]{Low Temperature Laboratory, Aalto University School of Science, POB 13500, FI-00076 AALTO, Finland}
\affiliation{RIKEN Center for Emergent Matter Science, Wako, Saitama 351-0198, Japan}

\author{Z. H. Peng}
\affiliation{RIKEN Center for Emergent Matter Science, Wako, Saitama 351-0198, Japan}

\author{Yu. P. Korneeva}
\affiliation{Moscow State Pedagogical University, 01069, Moscow, Russia}

\author{B. M. Voronov}
\affiliation{Moscow State Pedagogical University, 01069, Moscow, Russia}

\author{A. A. Korneev}
\affiliation{Moscow State Pedagogical University, 01069, Moscow, Russia}
\affiliation{Moscow Institute of Physics and Technology, 141700, Dolgoprudny, Moscow Region, Russia}
\affiliation{National Research University Higher School of Economics, Moscow Institute of Electronics and Mathematics, 109028, Moscow, Russia}

\author{A. V. Semenov}
\affiliation{Moscow State Pedagogical University, 01069, Moscow, Russia}
\affiliation{Moscow Institute of Physics and Technology, 141700, Dolgoprudny, Moscow Region, Russia}

\author{G. N. Gol'tsman}
\affiliation{Moscow State Pedagogical University, 01069, Moscow, Russia}
\affiliation{National Research University Higher School of Economics, Moscow Institute of Electronics and Mathematics, 109028, Moscow, Russia}

\author{J. S. Tsai}
\affiliation{RIKEN Center for Emergent Matter Science, Wako, Saitama 351-0198, Japan}
\affiliation{Department of Physics, Tokyo University of Science, Kagurazaka, Tokyo 162-8601, Japan}

\author{O. V. Astafiev}
\email{Oleg.Astafiev@rhul.ac.uk}
\affiliation{Royal Holloway, University of London, Egham, Surrey TW20 0EX, United Kingdom}
\affiliation{National Physical Laboratory, Hampton Road, Teddington TW11 0LW, United Kingdom}
\affiliation{RIKEN Center for Emergent Matter Science, Wako, Saitama 351-0198, Japan}
\affiliation{Moscow Institute of Physics and Technology, 141700, Dolgoprudny, Moscow Region, Russia}

\date{\today}

\begin{abstract}

We demonstrate coherent dynamics of quantized magnetic fluxes in a superconducting loop with a weak link -- a nanobridge patterned from the same thin NbN film as the loop. The bridge is a short rounded shape constriction, close to 10 nm long and 20 -- 30 nm wide, having minimal width at its center. Quantum state control and coherent oscillations in the driven time evolution of the tunnel-junctionless system are achieved. Decoherence and energy relaxation in the system are studied using a combination of microwave spectroscopy and direct time-domain techniques. The effective flux noise behavior suggests inductance fluctuations as a possible cause of the decoherence.

\end{abstract}

\maketitle

{\it Introduction.} A variety of different superconducting artificial quantum systems, successfully developed and studied over the last decade, rely on aluminum-based Josephson tunnel junctions of the Superconductor -- Insulator -- Superconductor (SIS) type (see, e.g.,~\cite{devoret13,paik11,barends14,chow14}). A recently proposed and explored alternative approach to superconducting qubits is based on a new phenomenon -- coherent quantum phase slips (CQPS) occurring in superconducting nanowires~\cite{mooij05,mooij06}. Such systems were first demonstrated in disordered $\inox$~\cite{astafiev12}, and we discovered similar behavior in nanowires patterned from thin disordered NbN and TiN films~\cite{peltonen13}.

{\it The main findings.}
The goal of this work is to realize a novel type of a tunnel barrier for magnetic fluxes based on the single weak link patterned from a disordered superconductor, and to study its coherent quantum dynamics, decoherence, and energy relaxation. In contrast to nanowire-based devices, the tunneling energy of the weak link is mainly determined by a single amplitude through the narrowest point. This helps to avoid interference between different amplitudes (Aharonov--Casher effect), which may take place in nanowire-based qubits, and energy fluctuations due to the random charge jumps in dielectrics~\cite{vanevic12}. Investigating the coherent dynamics in a weak link is interesting in light of the intense research on the classical dynamics and transport in these basic superconducting structures~\cite{likharev79}. Here we successfully demonstrate coherent flux tunneling dynamics through the weak link, together with quantum state control. Furthermore, we report a systematic investigation of decoherence and find constraints on the decoherence mechanisms in the system. Quantum dynamics in various types of superconducting weak links is under active study, and has been only recently observed in direct time-domain measurements for atomic contacts~\cite{janvier15,janvier14} and semiconducting nanowire-based SNS junctions~\cite{larsen15,delange15}.

\begin{figure}[!htb]
\includegraphics[width=\columnwidth]{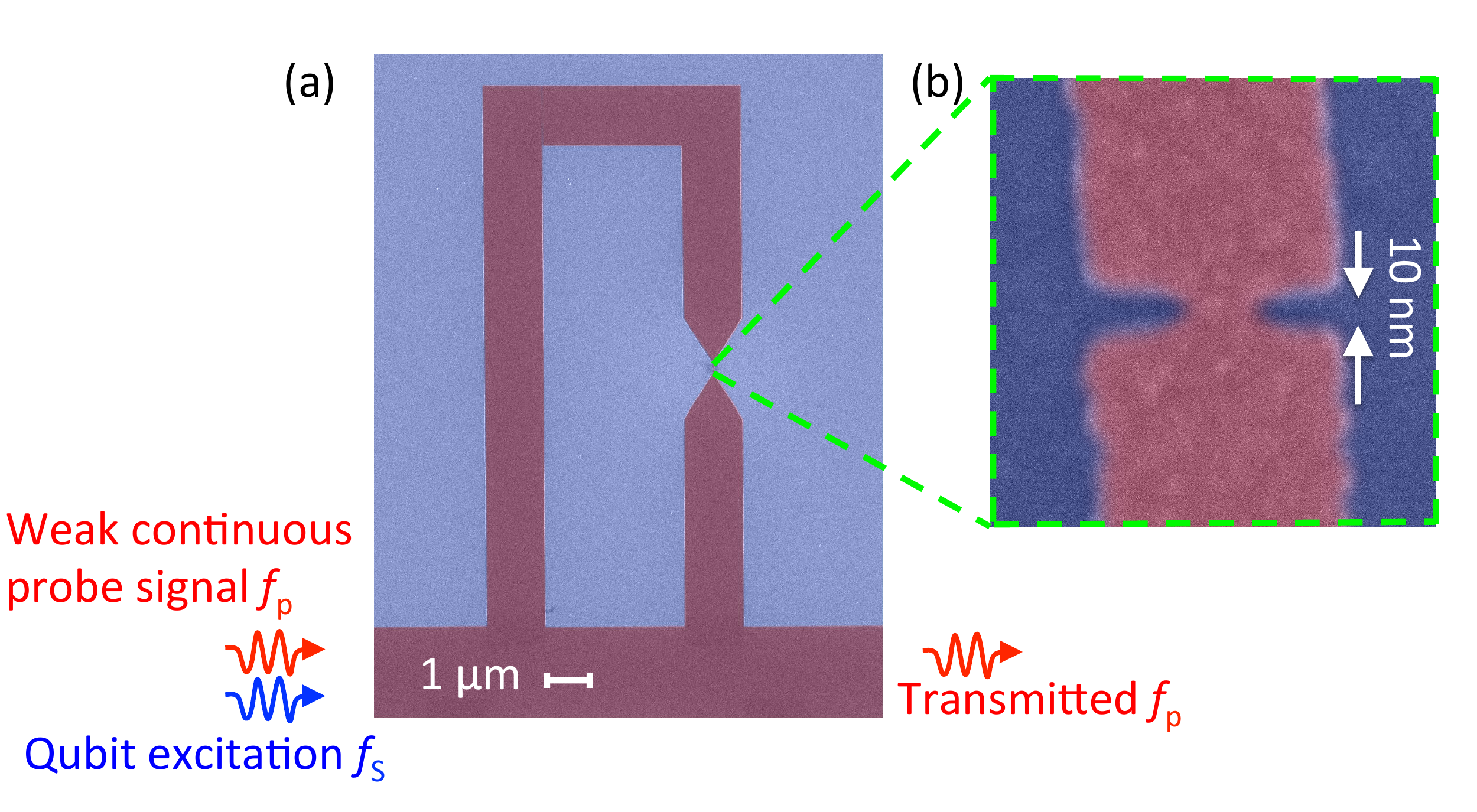}
\caption{(color online) \figta False color scanning electron micrograph of a typical NbN loop, interrupted by a narrow constriction. To control fluxes a perpendicular magnetic field $\bext$ is applied, producing a flux $\phiext$ through the loop. The bottom edge of the loop is shared with the center conductor of a NbN coplanar waveguide resonator for readout and control. \figtb Enlarged view of the Dayem bridge type constriction. } \label{fig:scheme}
\end{figure}

Figure~\ref{fig:scheme}~\figtb shows a closeup of a typical investigated constriction. During electron beam lithography as part of the fabrication process, the weak link is written as a short single pixel line between two $100\nm$ wide electrodes. The constriction lengths of the detected qubits were in the range $10-30\nm$, whereas the widths observed with a scanning electron microscope varied typically between 20 and $40\nm$. In this limit of short electrode separations, due to the fabrication process, the nanobridge length and width are not independent of each other: the width typically decreases with increasing length. Here we focus on one of the several measured chips, and present detailed measurements on one of the observed qubits, with the optimal point at $\fq=9.58\ghz$.

{\it Sample details.}
The samples are fabricated using a process similar to Ref.~\cite{peltonen13}: First, a NbN film of thickness $d\approx 2-3\nm$ is deposited on a Si substrate by DC reactive magnetron sputtering~\cite{goltsman03,korneev11}. Proceeding with the uniform NbN film, coplanar resonator groundplanes as well as the transmission lines for connecting to the external microwave measurement circuit are patterned in a first round of electron beam lithography (EBL) and subsequently metallized in an electron beam evaporator. In a second EBL step, the loops with constrictions as well as the resonator center line are patterned using a high resolution negative resist (calixarene)~\cite{tebn1,fujita96,narihiro05}. Reactive ion etching (RIE) in $\mr{CF}_4$ plasma is then used to transfer the pattern into the NbN film.

For electrical characterization of the samples, we use a weak continuous microwave of frequency $\fp$ as a probe signal and measure microwave transmission (normalized complex transmission coefficient $t=|t|e^{i\phi}$) through the resonator around one of the resonant modes using a vector network analyzer. The mode is chosen to be within the usable $6-12\ghz$ bandwidth of our cryogenic amplifier. We denote the probing power at the generator output by $\ppp$. The qubits can be excited using a second, continuous or pulsed, microwave tone at frequency $\fs$ and power $\pps$.

The sample reported here contains a resonator with capacitive coupling. The resonant modes are given by $f_n=n\vph/(2L)$, $n=1,2,3,\ldots$, where $L=1.5\mm$ is the resonator length, $\vph=1/(\llen\clen)^{1/2}$ the effective speed of propagation of the electromagnetic waves, and $\llen$ ($\clen$) the inductance (capacitance) per unit length. The qubit, shifted from the center of the resonator by about $100\mum$, is coupled to different resonator modes, depending on the oscillating current amplitude. Transmission measurements yielded an average mode spacing of $2\ghz$. We find $\vph\approx 6\times 10^{6}\ms$, and using the estimate $\clen\approx 1.0\times10^{-10}\fm$ obtain $\llen\approx 2.8\times 10^{-4}\hm$, corresponding to the square inductance $\lsq\approx 1.4\nh$ and characteristic impedance $Z_1=(\llen/\clen)^{1/2}\approx 1.7\kohm$. The full width at half maximum of the power transmission peak $\Delta f_6$ at $n=6$ with frequency $f_6=11.876\ghz$ (where the most of measurements are done) is about $23\mhz$, which corresponds to the quality factor  $Q \approx 500$. The resonator chip was enclosed in a sample box, and microwave characterization of the qubits is performed in a dilution refrigerator at the base temperature close to $25\mk$.

{\it Qubit characterization.}
\begin{figure}[!htb]
\includegraphics[width=\columnwidth]{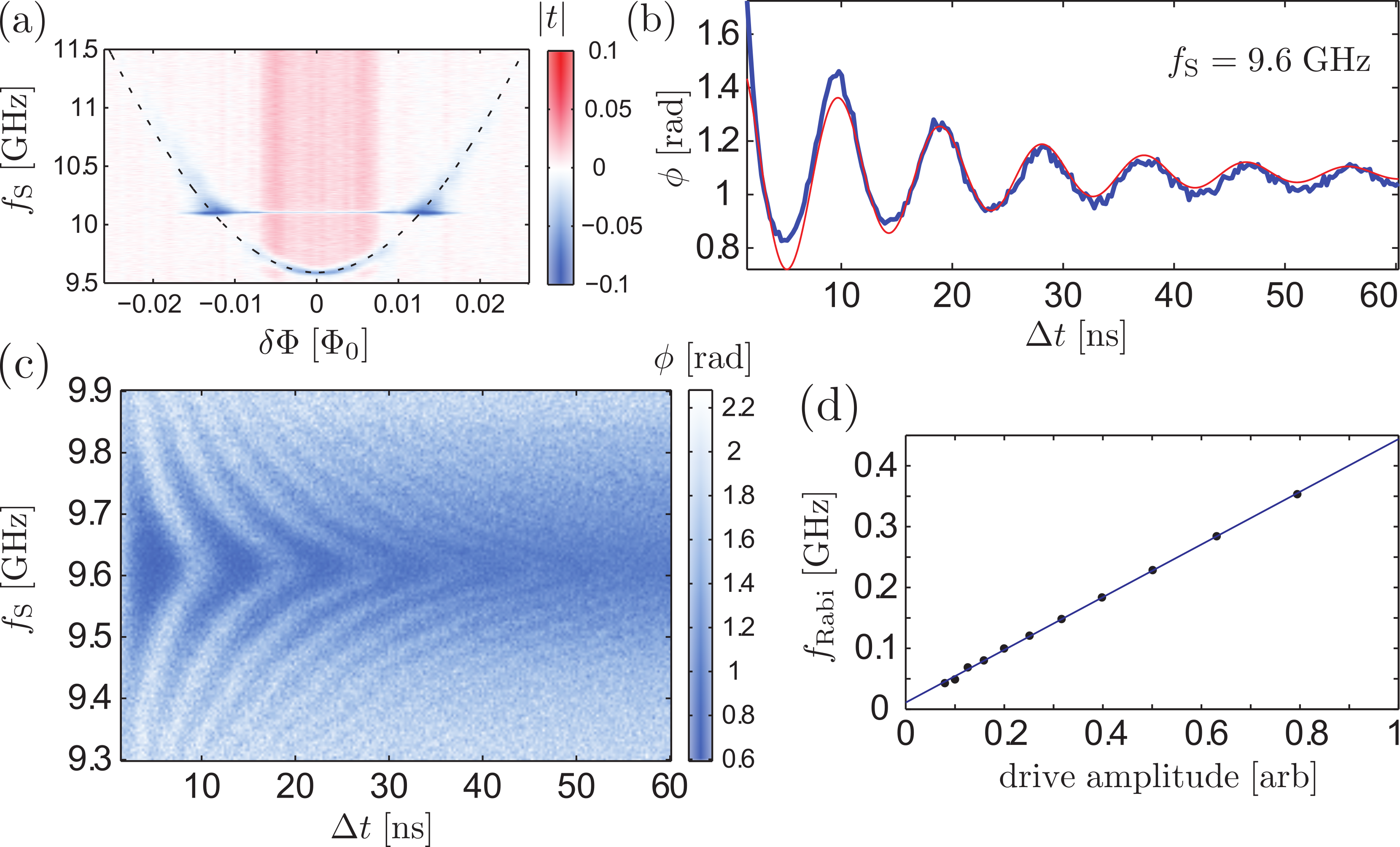}
\caption{(color online) \figta Two-tone spectroscopy of the system. The dashed line shows a fit with $\Delta/h=9.58\ghz$ and $\ip=40\na$. A $\fs$-independent background signal due to the change of $|t|$ with $\phiext$ (at $\fp$) has been subtracted. \figtb Coherent oscillations close to the optimal point ($\fs=\Delta/h$), measured by changing the pulse length $\Delta t$ of the pulsed, resonant microwave drive. The shown phase is the measured average phase when the system is continuously probed and the excitation pulse is repeated at intervals of $T=500\ns$ ($\gg T_1$, the energy relaxation time). The thin red line shows a fit to a decaying sinusoidal. \figtc Rabi oscillations close to the optimal point, for a range of driving frequencies $\fs$. The trace in panel \figtb corresponds to a line cut of a similar plot. \figtd Dependence of Rabi frequency on the driving amplitude, showing linear increase towards stronger driving as evidence of the two-level nature of the system.} \label{fig:rabi}
\end{figure}
We now demonstrate that the weak links work as a tunnel barrier for magnetic fluxes. Microwave characterization of the resonators typically showed flux-periodic signatures originating from several of the loops coupled to the same readout resonator. At first, to identify which of the several loops, coupled to the center line of the shared resonator, form functioning and detectable qubits, we probe the microwave transmission around several resonator modes as a function of $\phiext$ over several flux periods. As in Ref.~\onlinecite{peltonen13}, the loop area increases by a factor of 3 between the smallest and the largest loop. The initial identification based on the flux-periodicity of the transmission is followed by two-tone spectroscopy centered around the optimal point of each qubit to extract the minimum energy gap $\Delta$ (the magnetic flux tunneling energy) and the persistent current $\ip$ from the $\phiext$-dependence of the qubit transition. Figure~\ref{fig:rabi}~\figta shows a typical result of such a measurement: Microwave transmission through the resonator is monitored continuously at a fixed frequency $\fp$ using a weak probe tone at one of the resonant modes. The qubit is simultaneously excited using a stronger drive tone at a frequency $\fs$ that is scanned around $\fq$. Due to the dispersive (non-resonant) coupling between the qubit and the resonator, the resonant frequency depends on the qubit state populations. The populations ultimately saturate to 0.5 for strong drive, producing a clear dip in the measured transmission whenever $\fs=\fq$. In Fig.~\ref{fig:rabi}~\figta we plot the magnitude of the microwave transmission coefficient as a function of $\fs$ and $\phiext$. The dashed line is a fit to $h\fq=\sqrt{\varepsilon^2+\Delta^2}$, based on the Hamiltonian $H=-(\varepsilon/2)\sigma_z-(\Delta/2)\sigma_x$. Here, $\varepsilon=2\ip\dphi$ with $\dphi=\phiext-(N+1/2)\phio$, and $\phio=h/(2e)$ denotes the quantum of magnetic flux. In the vicinity of the point $\dphi=0$, where the transition frequency is minimal reaching $\Delta/h$, the fluxes are superposed. The dashed curve uses $\ip=40\na$, which is defined by the loop inductance $\lq=\Phi_0/(2\ip)\approx 25\nh$.

After spectroscopic characterization, we can perform direct time-domain probing with $\phiext$ tuned to the optimal point, where $\fq=\Delta/h$. Fig.~\ref{fig:rabi}~\figtc shows the second main finding of this Letter: Using a pulsed microwave drive at frequency $\fs$ close to $\fq$, and of varying duration $\Delta t$, we observe Rabi oscillations of the qubit population. To obtain this plot, we keep the weak continuous probing tone at $\fp$, whereas the driving pulses of length $\Delta t$ are repeated with period $T=500\ns$. Figure~\ref{fig:rabi}~\figtb shows a line cut at zero detuning, $\fs=\fq$, whereas Fig.~\ref{fig:rabi}~\figtd illustrates good agreement between the observed oscillation frequencies at different driving powers and the characteristic linear dependence on the drive amplitude expected for a two level system. The oscillations decay here within about $30\ns$. In addition to the decay, we also observe a slow rise or fall of the background level, without clear nature. One of the possible causes is influence of the excitation pulses on the resonator transmission.

The time-domain oscillation measurements were performed at probing powers $\ppp$ corresponding to intraresonator photon numbers $n\lesssim 1$. To characterize the influence of the continuous measurement on the qubit dephasing we check the lineshape of the spectroscopy line at the qubit optimal point at varying probing powers. The observed negative ac-Stark shift of the qubit frequency, approximately linear with an increase of the readout power, allowed us to calibrate $n$ against $\ppp$ as with aluminum-based superconducting qubits~\cite{schuster05,abdumalikov08}. Due to the measurement backaction, at increasing $\ppp$ we observe the expected broadening of the spectroscopy line and the start of the evolution of its shape from a Lorentzian towards a Gaussian.

\begin{figure}[!htb]
\includegraphics[width=\columnwidth]{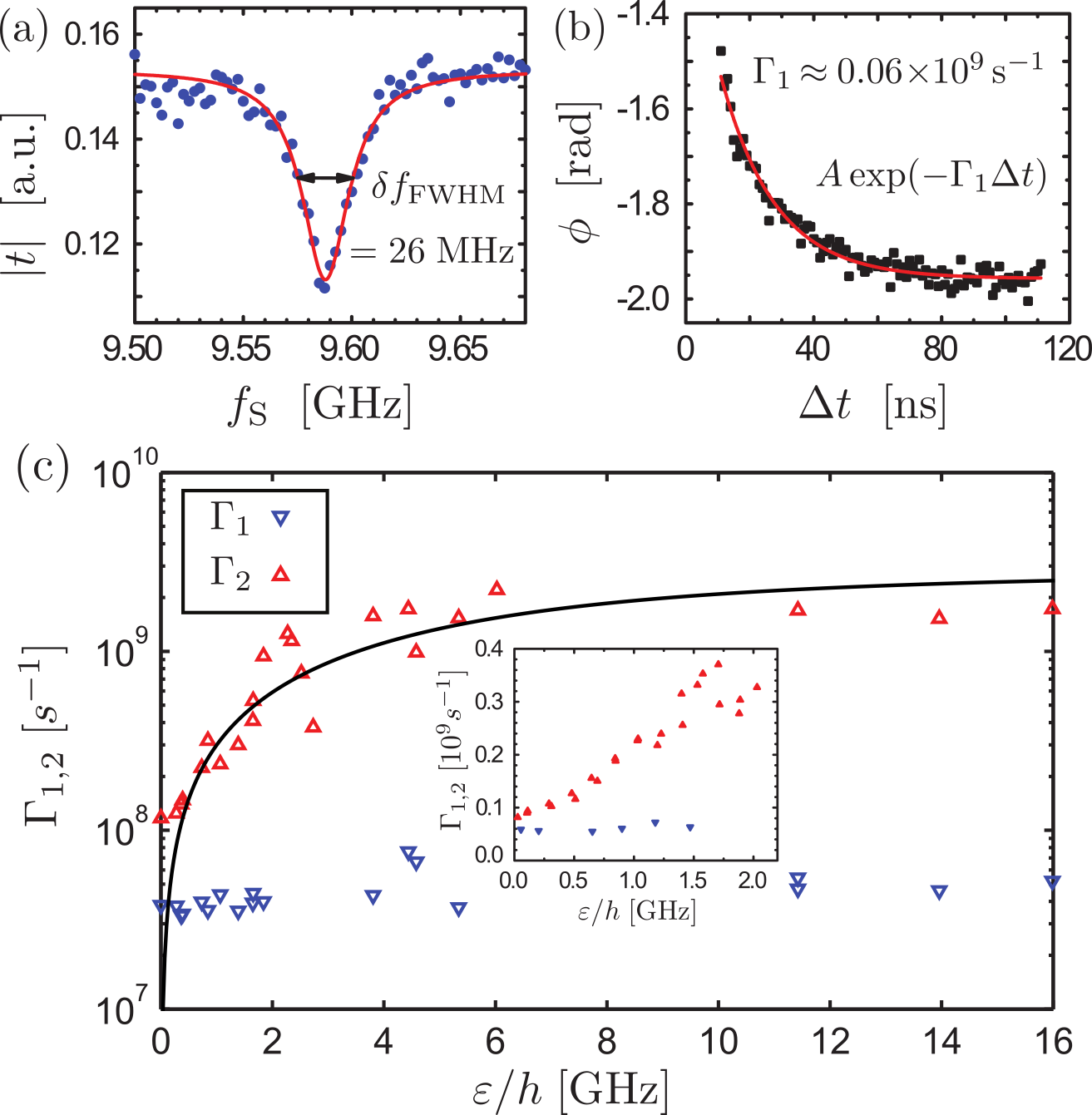}
\caption{(color online) \figta Typical spectroscopy lineshape of the qubit close to the optimal point (filled circles). The probing power $\ppp$ was kept constant at a low value corresponding to $n\lesssim1$ photons in the readout resonator, in which case the measured signal is well described by a Lorentzian (red solid line), the width of which relates directly to the total decoherence rate $\Gamma_2=\pi\deltaff$. \figtb Determination of the relaxation rate $\Gamma_1$ from exponential decay of the averaged transmission under repetitive driving with double pulses of constant length but varying delay $\Delta t$ between the pulses. \figtc Relaxation rates $\Gamma_1$ (downward blue/dark triangles) and total decoherence rates $\Gamma_2$ (upward red/light triangles) as a function of the energy bias $\varepsilon$. The solid line shows $2.9|\varepsilon|/(\hbar\wq)\times 10^{9}\pers$. The inset plots $\Gamma_1$ and $\Gamma_2$ from another, more detailed measurement focused around the optimal point.} \label{fig:decoherence}
\end{figure}

{\it Decoherence measurements.}
Our earlier experiments with two-level systems in $\inox$ and NbN nanowires~\cite{astafiev12,peltonen13} indicated dephasing in these systems, with typical spectroscopy linewidths of the order of $100\mhz$, but the decoherence properties were not investigated in detail. Figure~\ref{fig:decoherence} collects together the next finding of this Letter: measurements of dephasing and relaxation rates of the constriction qubit.

First, Fig.~\ref{fig:decoherence}~\figta displays the spectroscopy line of the qubit when $\phiext$ is kept fixed at the optimal point. The probing power $\ppp$ corresponds to $n\lesssim1$ photons in the resonator on average, and the magnitude of the normalized transmission coefficient is plotted as a function of $\fs$. The lineshape is well described by a Lorentzian dip whose width and depth increase with $\pps$. In the limit of low $\pps$ we find a full width at half maximum of $\deltaff = 26\mhz$ at the optimal point, corresponding to $\Gamma_2\approx \pi\deltaff\approx 8\times 10^{7}\pers$.

The dependence of the dephasing rate $\Gamma_2$ on the qubit frequency $\fq$ is shown in Fig.~\ref{fig:decoherence}~\figc, on the main panel in terms of $\varepsilon$ over a wide range of the external flux, and in the inset in more detail around the optimal point. As evident already from Fig.~\ref{fig:rabi}~\figa, the dephasing rates increase quickly when $\phiext$ is moved away from the optimal point. The black solid line as an eye-guide is $2.9|\varepsilon|/(\hbar\wq)\times 10^{9}\pers$, the expected $\varepsilon$-dependence for $1/f$ type flux noise~\cite{yoshihara06}. In Fig.~\ref{fig:decoherence}~\figtc we also show the approximate energy relaxation rates $\Gamma_1=1/T_1$ as a function of the external magnetic flux. To reduce the effects of spurious excitations of nearby resonator modes, the qubit relaxation rates were obtained from exponential fits to a time domain measurement with two microwave pulses of fixed length $100\ns\gg T_2$ whose separation was varied. Figure~\ref{fig:decoherence}~\figtb displays a typical result. Similar to the measurements of the Rabi oscillations, the system was probed continuously at $\fp$ and the pulse sequence was repeated after every $T=500\ns$. In contrast to the dephasing rate that quickly increases away from the qubit optimal point, we find the relaxation rate corresponding to $T_1\approx 30\ns$ to be weakly dependent of $\phiext$. Note also that the relaxation can not be described by the Purcell effect due to non-resonant energy leak to the resonator modes. This is supported by our estimates and frequency-independent behavior of $\Gamma_1$ when the qubit is detuned from the resonances, as is, for instance, shown in the inset of Fig.~\ref{fig:decoherence}~\figc.

{\it Discussions.}
The dephasing dependence in Fig.~\ref{fig:decoherence}~\figtc suggests that it originates from low frequency fluctuations in the flux degree of freedom. From the asymptotic relation $2 \ip \delta \Phi = 2 \hbar \Gamma_2$ (at $\hbar\wq \gg \Delta$, where $\Gamma_2 \approx 2.9 \times 10^9\pers$), we find the corresponding normalised flux fluctuations $\delta\Phi/\Phi_0$ to be about $4\times 10^{-3}$, which is about three orders of magnitude larger than typical flux fluctuations in Josephson flux qubits and dc SQUIDs fabricated from thicker films~\cite{yoshihara06,yan12,stern14,orgiazzi14,anton13,yan15}. The corresponding inductance fluctuations from $\delta\Phi/\Phi_0 = \delta\lq/\lq$ are found to be $\delta\lq \approx 0.1\nh$. If we assume that the fluctuations are correlated in space with a typical characteristic length (e.g. coherence length) then the relative inductance fluctuations $\delta\lq/\lq$ are scaled as inverse square root of the total area.
Recalculated for the resonator, we find that $\delta\lr/\lr= \sqrt{\aq/\ar}\delta\lq/\lq \approx 2.5\times 10^{-4}$, where $\aq$ and $\ar$ are areas of the qubit and the resonator central line. If similar inductance fluctuations take place in the resonator formed from the same NbN film, the expected relative inductance fluctuations result in the resonator line broadening $\delta f = 1/2 (f_n/2) \delta \lr/\lr$, which for $f_6$ is equal to $0.8\mhz$. This is an order of magnitude smaller than the line broadening in our resonator $\Delta f_6$.

Next, we characterize the qubit dissipation, assuming that it is caused by a resistance $\rrq$ parallel to the inductance. The relaxation rate caused by the spontaneous emission due to the current quantum noise $S_I(\wq) = 2\hbar\wq/(2\pi\rrq)$ is $\Gamma_1 = 2\pi S_I(\wq) \phip^2 \sin^2\theta/\hbar^2$, where $\phip = \lq \ip = \Phi_0/2$ is the effective dipole moment, describing coupling of the loop to the resistance, and $\sin\theta = \Delta/\hbar\wq$. Note that $\sin\theta$ changes only by a factor of 2 in the range of our measurements $9.58\ghz\leq\fq\leq 18\ghz$ (corresponding to $0\leq\epsilon\leq 16\ghz)$, and $\Gamma_1$ fluctuates in a relatively narrow window over the measured frequency range. Although $\rrq$ is not necessarily constant, we estimate its effective value, substituting typical $\Gamma_1$. At $\fq\approx 9.6\ghz$, $\Gamma_1 \approx 4 \times 10^{7}\pers$, and we find $\rrq\approx 30\Mohm$, which corresponds to a square resistance $\rsq\approx 1.7\Mohm$. The nature of this dissipation requires further study. A possible mechanism can be related to crowding of the current part out of superconducting channel to a normal one (e.g., quasiparticle current or dissipative displacement current in the oxide layer on top of the film) due to fluctuations of kinetic inductance.

{\it Conclusions.}
To summarize, we have demonstrated coherent quantum dynamics in a superconducting loop interrupted by a weak link in the geometry of a uniform-thickness Dayem bridge type constriction. Quantum state control of the qubit has been demonstrated by measuring Rabi oscillations. The dephasing and energy relaxation has been studied in a wide range of energies. The dephasing can be explained by kinetic inductance fluctuations in the highly disordered NbN film. Future samples would benefit from a readout resonator with significantly larger fundamental frequency, better quality resonators fabricated in a separate step, as well as the loops fabricated from a thicker film. A detailed study of the constriction length dependence could shed light onto the transition to a phase slip flux qubit in a longer nanowire, either uniform or behaving as a chain of intrinsic weak links.

\begin{acknowledgments}
The work was financially supported by the JSPS FIRST and ImPACT programs, MEXT Kakenhi 'Quantum Cybernetics', Royal Society within International Exchanges Scheme (2015 Russia RFBR \#15-52-10044 KO\_a Cost share), and Russian Scientific Fund (N 15-12-30030). J. T. P. acknowledges support from Academy of Finland (Contract No. 275167).
\end{acknowledgments}


\begin{thebibliography}{99}

\bibitem{devoret13} M. H. Devoret and R. J. Schoelkopf, Science {\bf 339}, 1169 (2013).

\bibitem{paik11} H. Paik, D. I. Schuster, Lev S. Bishop, G. Kirchmair, G. Catelani, A. P. Sears, B. R. Johnson, M. J. Reagor, L. Frunzio, L. I. Glazman, S. M. Girvin, M. H. Devoret, and R. J. Schoelkopf, Phys. Rev. Lett. {\bf 107}, 240501 (2011).

\bibitem{barends14} R. Barends {\it et al.}, Nature {\bf 508}, 500 (2014).

\bibitem{chow14} J. M. Chow, J. M. Gambetta, E. Magesan, S. J. Srinivasan, A. W. Cross, D. W. Abraham, N. A. Masluk, B. R. Johnson, C. A. Ryan, and M. Steffen, Nat. Comm. {\bf 5}, 4015 (2014).

\bibitem{mooij05} J. E. Mooij and C. J. P. M. Harmans, N. J. Phys. {\bf 7}, 219 (2005).

\bibitem{mooij06} J. E. Mooij and Yu. V. Nazarov, Nat. Phys. {\bf 2}, 169 (2006).

\bibitem{astafiev12} O. V. Astafiev, L. B. Ioffe, S. Kafanov, Yu. A. Pashkin, K. Yu. Arutyunov, D. Shahar, O. Cohen, and J. S. Tsai, Nature {\bf 484}, 355 (2012).

\bibitem{peltonen13} J. T. Peltonen, O. V. Astafiev, Yu. P. Korneeva, B. M. Voronov, A. A. Korneev, I. M. Charaev, A. V. Semenov, G. N. Gol'tsman, L. B. Ioffe, T. M. Klapwijk, and J. S. Tsai, Phys. Rev. B {\bf 88}, 220506(R) (2013).

\bibitem{vanevic12} M. Vanevic and Yu. V. Nazarov, Phys. Rev. Lett. {\bf 108}, 187002 (2012).

\bibitem{likharev79} K. K. Likharev, Rev. Mod. Phys. {\bf 51}, 101 (1979).

\bibitem{janvier15} C. Janvier, L. Tosi, L. Bretheau, \c C. \"O. Girit, M. Stern, P. Bertet, P. Joyez, D. Vion, D. Esteve, M. F. Goffman, H. Pothier, and C. Urbina, Science {\bf 349}, 1199 (2015).

\bibitem{janvier14} C. Janvier, L. Tosi, \c C. \"O. Girit, M. F. Goffman, H. Pothier, and C. Urbina, J. Phys.: Condens. Matter {\bf 26}, 474208 (2014).

\bibitem{larsen15} T. W. Larsen, K. D. Petersson, F. Kuemmeth, T. S. Jespersen, P. Krogstrup, J. Nygard, C. M. Marcus, Phys. Rev. Lett. {\bf 115}, 127001 (2015).

\bibitem{delange15} G. de Lange, B. van Heck, A. Bruno, D. J. van Woerkom, A. Geresdi, S. R. Plissard, E. P. A. M. Bakkers, A. R. Akhmerov, and L. DiCarlo, Phys. Rev. Lett. {\bf 115}, 127002 (2015).

\bibitem{goltsman03} G. N. Gol'tsman, K. Smirnov, P. Kouminov, B. Voronov, N. Kaurova, V. Drakinsky, J. Zhang, A. Verevkin, and R. Sobolewski, IEEE Trans. Appl. Supercond. {\bf 13}, 192 (2003).

\bibitem{korneev11} A. Korneev, Yu. Korneeva, I. Florya, B. Voronov, and G. Gol'tsman, Proc. SPIE {\bf 8072}, 80720G (2011).

\bibitem{tebn1} Commercially available as 'TEBN-1' from Tokuyama Corporation, www.tokuyama.co.jp.

\bibitem{fujita96} J. Fujita, Y. Ohnishi, Y. Ochiai, and S. Matsui, Appl. Phys. Lett. {\bf 68}, 1297 (1996).

\bibitem{narihiro05} M. Narihiro, K. Arai, M. Ishida, Y. Ochiai, and Y. Natsuka, Jpn. J. Appl. Phys. {\bf 44}, 5581 (2005).

\bibitem{astafiev10} O. V. Astafiev, A. A. Abdumalikov, Jr., A. M. Zagoskin, Yu. A. Pashkin, Y. Nakamura, and J. S. Tsai, Phys. Rev. Lett. {\bf 104}, 183603 (2010).

\bibitem{abdumalikov11} A. A. Abdumalikov, Jr., O. V. Astafiev, Yu. A. Pashkin, Y. Nakamura, and J. S. Tsai, Phys. Rev. Lett. {\bf 107}, 043604 (2011).

\bibitem{schuster05} D. I. Schuster, A. Wallraff, A. Blais, L. Frunzio, R.-S. Huang, J. Majer, S. M. Girvin, and R. J. Schoelkopf, Phys. Rev. Lett. {\bf 94}, 123602 (2005).

\bibitem{abdumalikov08} A. A. Abdumalikov, Jr., O. V. Astafiev, Y. Nakamura, Y. Pashkin, and J. S. Tsai, Phys. Rev. B {\bf 78}, 180502(R) (2008).

\bibitem{yoshihara06} F. Yoshihara, K. Harrabi, A. O. Niskanen, Y. Nakamura, and J. S. Tsai, Phys. Rev. Lett. {\bf 97}, 167001 (2006).

\bibitem{yan12} F. Yan, J. Bylander, S. Gustavsson, F. Yoshihara, D. G. Cory, Y. Nakamura, and W. D. Oliver, Phys. Rev. B {\bf 85}, 174521 (2012).

\bibitem{stern14} M. Stern, G. Catelani, Y. Kubo, C. Grezes, A. Bienfait, D. Vion, D. Esteve, and P. Bertet, Phys. Rev. Lett. {\bf 113}, 123601 (2014).

\bibitem{orgiazzi14} J.-L. Orgiazzi, C. Deng, D. Layden, R. Marchildon, F. Kitapli, F. Shen, M. Bal, F. R. Ong, and A. Lupascu, arXiv:1407.1346 (2014).

\bibitem{anton13} S. M. Anton, J. S. Birenbaum, S. R. O'Kelley, V. Bolkhovsky, D. A. Braje, G. Fitch, M. Neeley, G. C. Hilton, H.-M. Cho, K. D. Irwin, F. C. Wellstood, W. D. Oliver, A. Shnirman, and J. Clarke, Phys. Rev. Lett. {\bf 110}, 147002 (2013).

\bibitem{yan15} F. Yan, S. Gustavsson, A. Kamal, J. Birenbaum, A. P. Sears, D. Hover, T. J. Gudmundsen, J. L. Yoder, T. P. Orlando, J. Clarke, A. J. Kerman, and W. D. Oliver, arXiv:1508.06299 (2015).

\end{thebibliography}
\end{document}